\documentclass[conference]{IEEEtran}
\IEEEoverridecommandlockouts

\usepackage{cite}
\usepackage{amsmath,amssymb,amsfonts}
\usepackage{graphicx}
\usepackage{textcomp}
\usepackage{subfig}
\usepackage{algorithm}
\usepackage{algorithmicx, program, algpseudocode}
\usepackage{authblk}

\usepackage{color}
\definecolor{RED1}{rgb}{1,0,0}
\definecolor{BLUE1}{rgb}{0,0,1}

\newcommand{\etal}{\textit{et al}.}
\newcommand{\ie}{\textit{i}.\textit{e}.}
\newcommand{\eg}{\textit{e}.\textit{g}.}

\ifCLASSINFOpdf
\else
\fi
\begin{document}
%
\title{Bilateral-ViT for Robust Fovea Localization}

\author[1 \IEEEauthorrefmark{1}]{Sifan Song\thanks{* Contributed Equally}}
\author[2 \IEEEauthorrefmark{1}]{Kang Dang}
\author[3]{Qinji Yu}
\author[2]{Zilong Wang}
\author[4]{Frans Coenen}
\author[1 \IEEEauthorrefmark{2}]{Jionglong Su}
\author[2,3 \IEEEauthorrefmark{2}]{Xiaowei Ding\thanks{\IEEEauthorrefmark{2} Corresponding Authors \par E-mail: Jionglong.Su@xjtlu.edu.cn (J. Su); \par \,\,\,\,\,\,\,\,\,\,\,\,\,\,\,\,\,\, dingxiaowei@sjtu.edu.cn (X. Ding) \par
This work has been accepted for oral presentation by ISBI2022.}}
\affil[1]{\small Xi'an Jiaotong-Liverpool University, Suzhou, China}
\affil[2]{VoxelCloud, Inc., Los Angeles, USA}
\affil[3]{Shanghai Jiao Tong University, Shanghai, China}
\affil[4]{University of Liverpool, Liverpool, UK}


\maketitle

\begin{abstract}
The fovea is an important anatomical landmark of the retina. Detecting the location of the fovea is essential for the analysis of many retinal diseases.  However, robust fovea localization remains a challenging problem, as the fovea region often appears fuzzy, and retina diseases may further obscure its appearance. This paper proposes a novel Vision Transformer (ViT) approach that integrates information both inside and outside the fovea region to achieve robust fovea localization. Our proposed network, named Bilateral-Vision-Transformer (Bilateral-ViT), consists of two network branches: a transformer-based main network branch for integrating global context across the entire fundus image and a vessel branch for explicitly incorporating the structure of blood vessels. The encoded features from both network branches are subsequently merged with a customized Multi-scale Feature Fusion (MFF) module. Our comprehensive experiments demonstrate that the proposed approach is significantly more robust for diseased images and establishes the new state of the arts using the \texttt{Messidor} and \texttt{PALM} datasets.
\end{abstract}


%
\IEEEpeerreviewmaketitle

\section{Introduction}
\label{sec:intro}

The macula is the central region of the retina. The fovea is an important anatomical landmark located in the center of the macula, responsible for the most crucial part of a person's vision~\cite{weiter1984visual}. The severity of vision loss due to retinal diseases is usually related to the distance between the associated lesions and the fovea. Therefore, detecting the location of the fovea is essential for the analysis of many retinal diseases. 

Despite its importance, robust fovea localization remains a challenging problem.  The color contrast between the fovea region and its surrounding tissue is poor, leading to a fuzzy appearance. Furthermore, the fovea appearance may be obscured by lesions in the diseased retina; for example, geographic atrophy and hemorrhages significantly alter the fovea appearance. Such issues make it more difficult to perform localization based on the fovea appearance alone. Fortunately, anatomical structures outside the fovea region, such as blood vessels, are also helpful for localization~\cite{li2004automated,aquino2014establishing}. For this reason, we propose a novel Vision Transformer (ViT) approach that integrates information both inside and outside the fovea region to achieve robust fovea localization.

Our proposed network, named Bilateral-Vision-Transformer (Bilateral-ViT), consists of two network branches.  We adopt a transformer-based U-net architecture~\cite{chen2021transunet} as the \textbf{main branch} for effectively integrating global context across the entire fundus image. In addition, we design a \textbf{vessel branch} that takes in a blood vessel segmentation map for explicitly incorporating the structure of blood vessels.  Finally, the encoded features from both network branches are merged with a customized Multi-scale Feature Fusion (MFF) module, leading to significantly improved performance. Thus, our key contributions are as follows: 
\begin{itemize}
	\item We propose a novel vision-transformer-based network architecture, that explicitly incorporates global image context and structure of blood vessels, for robust foveal localization. 
	\item We demonstrate that the proposed approach is significantly more robust for challenging settings such as fovea localization in diseased retinas (over 9\% improvements for specific evaluations). It also has a better generalization capability compared to the baseline methods, as shown in cross-dataset experiments.
	\item We establish the new state of the arts on both the \texttt{Messidor} and \texttt{PALM} datasets.
\end{itemize}

\section{Related Work}

Earlier work usually utilize hand-craft features to encode anatomical relationships among optic discs (OD), blood vessels, and fovea regions for fovea localization. Deka \etal~\cite{deka2015detection} and Medhi \etal~\cite{medhi2016effective} generate the region of interest (ROI) using processed blood vessels for macula estimation. Certain methods utilize OD in the prediction of ROI and fovea center by selecting specific OD diameters~\cite{narasimha2006robust}, estimating OD orientations and minimum intensity values~\cite{sekhar2008automated, asim2012detection}. Other applications use combined OD and blood vessels features to improve the performance of fovea localization~\cite{li2004automated,aquino2014establishing}. These methods generally perform less competitively than more recent deep-learning-based approaches.

Many deep learning-based methods formulate the fovea localization as a regression task~\cite{al2018multiscale,meyer2018pixel,huang2020efficient,xie2020end}. Some methods utilize retinal structures, such as OD and blood vessels, as constraints for inferring the location of the fovea. For example, Meyer \etal~\cite{meyer2018pixel} adopt a pixel-wise distance regression approach for joint OD and fovea localization. Besides the regression-based approaches, Sedai \etal~\cite{sedai2017multi} propose a two-stage image segmentation framework for segmenting the image region around the fovea. Our work also belongs to the image segmentation paradigm~\cite{chen2021transunet,sedai2017multi,ronneberger2015u,qin2020u2,yu2021location}. Unlike all previous works, we customize the recent transformer-based segmentation network~\cite{chen2021transunet} to incorporate blood vessel information and demonstrate its superior performance compared to the existing approaches.

\section{Methodology}
\label{sec:meth}
\subsection{Network Architecture}
\begin{figure}
	\centering
	\includegraphics[width=3in]{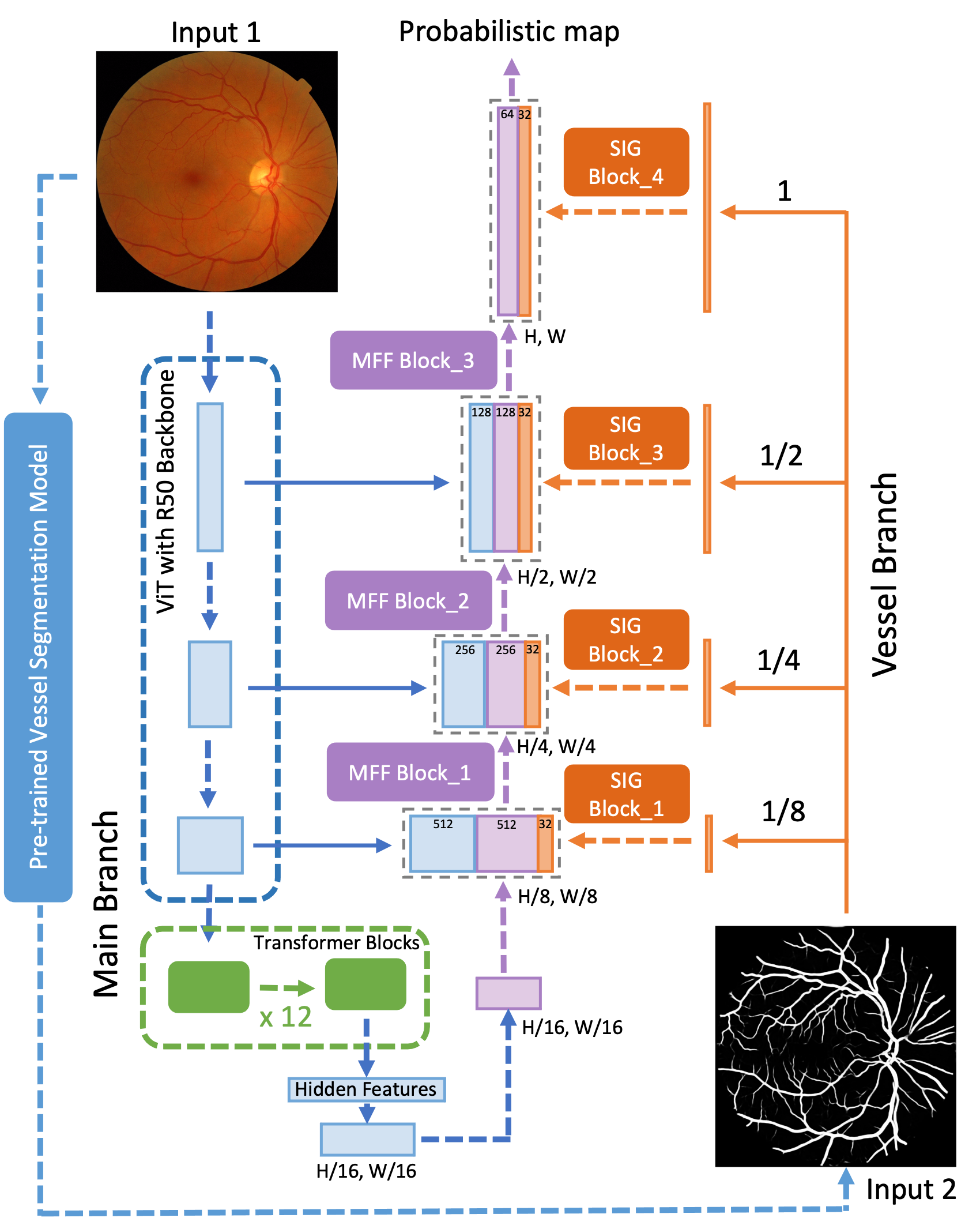}
	\caption{\footnotesize The overall architecture of our proposed Bilateral-ViT network.} \label{Fig_Arch1}
\end{figure}

The overall architecture of Bilateral-ViT is illustrated in Fig.~\ref{Fig_Arch1}. The proposed Bilateral-ViT is based on a U-shape architecture with a vision transformer-based encoder (\textbf{the main branch}) for exploiting long-range contexts.  In addition, we design a \textbf{vessel branch} to encode structure information from blood vessel segmentation maps.  Finally, Multi-scale Feature Fusion (MFF) blocks are designed to effectively fuse data from the main and vessel branches. 

\textbf{Main Branch.} We adopt the TransUNet~\cite{chen2021transunet} as the main branch due to its superior performance on other medical image segmentation tasks. In the main branch, we utilize a CNN-Transformer hybrid structure as the encoder. The CNN part is used as the initial feature extractor. It provides features at different scales for the skip connections to compensate for the information loss in the downsampling operation. The extracted features are then processed by 12 consecutive transformer blocks at the bottleneck of the UNet architecture. The transformer encodes the long-range dependencies of the input fundus image due to the multi-head self-attention structure. The output features of the last transformer block are then resized for later decoding operations.

\begin{figure}
	\centering
	\includegraphics[width=2.8in]{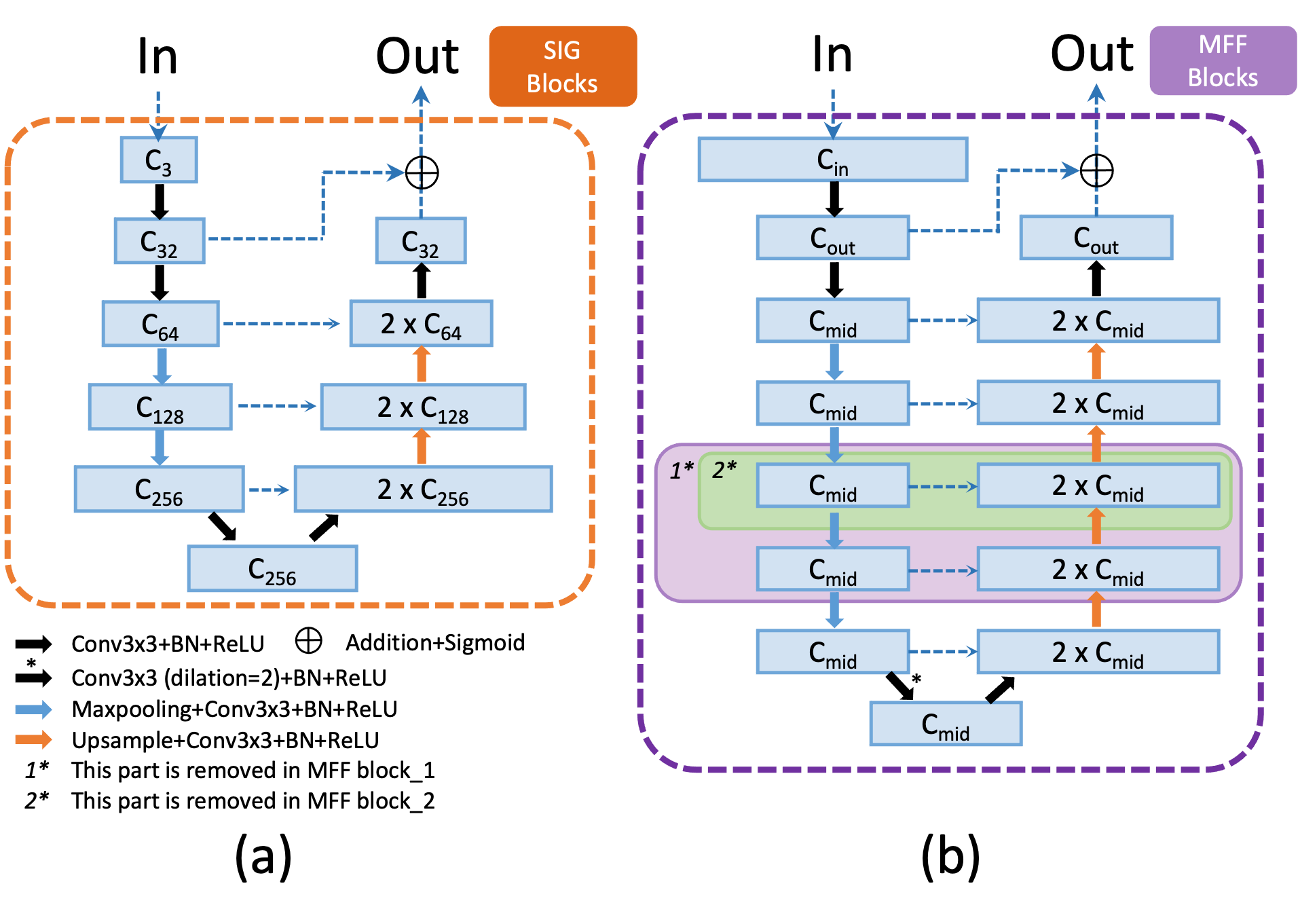}
	\caption{\footnotesize The structures of SIG blocks and MFF blocks.  The subscript $C$ denotes channel depths. $C_{in}$, $C_{mid}$ and $C_{out}$ represent channel depths of input, intermediate, and output feature maps for the MFF blocks, respectively. We set $C_{mid}$ of three MFF blocks to small numbers, \ie 128, 64, 32, for improving the efficiency of multi-scale feature fusion.} \label{Fig_Arch2}
\end{figure}

\textbf{Vessel Branch.} In the vessel branch, we aim to exploit the structure information from the blood vessels. Unlike the main branch, where the input is a fundus image, we put in a vessel segmentation map generated by a pre-trained model. The pre-trained vessel segmentation model is built on the DRIVE dataset~\cite{staal2004ridge} with the TransUNet~\cite{chen2021transunet} architecture. Four identical Spatial Information Guidance (SIG) blocks are utilized in the vessel branch to extract multi-scale vessel-based features. The rescaled vessel segmentation maps are fed into the SIG blocks, the details of which are illustrated in Fig.~\ref{Fig_Arch2}-a. The design of the SIG blocks makes extensive use of customized ReSidual U-blocks (RSU). Qin \etal~\cite{qin2020u2} indicate that the RSU block is superior in performance to other embedded structures (\eg, plain convolution, residual-like, inception-like, and dense-like blocks), due to the enlarged receptive fields of the embedded U-shape architecture.

\begin{table*}\scriptsize
	\caption{\footnotesize Comparison of performance on normal and diseased retinal images using the \texttt{Messidor} and \texttt{PALM} datasets. The best and second best results are highlighted in bold and italics respectively. }\label{table-path}
	\centering
	\begin{tabular}{|c|c|c|c|c|c|c|c|c|c|c|}
		\hline
		& \multicolumn{2}{|c}{1/8 R($\%$)} & \multicolumn{2}{|c}{1/4 R($\%$)} & \multicolumn{2}{|c}{1/2 R($\%$)} &  \multicolumn{2}{|c}{1R($\%$)} & \multicolumn{2}{|c|}{2R($\%$)} \\
		\hline
		\texttt{Messidor} & Normal & Diseased & Normal & Diseased & Normal & Diseased & Normal & Diseased & Normal & Diseased \\
		\hline
		UNet (2015)~\cite{ronneberger2015u} & 82.65 &	79.00 &	95.15 &	93.33 &	97.76 &	95.00 &	97.95 &	95.33 &	97.95 &	95.33\\
		U2 Net (2020)~\cite{qin2020u2} & 86.19 &	81.33 &	\textbf{98.51} &	97.33 &	\textit{99.63} &	99.50 &	\textit{99.63} &	99.50 &	\textit{99.63} &	99.50 \\
		TransUNet (2021)~\cite{chen2021transunet} & \textit{87.31} &	\textbf{84.33} &	\textit{98.32} &	\textit{97.67} &	\textbf{100.00} &	\textit{99.83} &	\textbf{100.00} &	\textit{99.83} &	\textbf{100.00} &	\textit{99.83} \\
		Bilateral-ViT (\textbf{Proposed}) & \textbf{87.50} &	\textit{84.00} &	\textbf{98.51} &	\textbf{98.67} &	\textbf{100.00} &	\textbf{100.00} &	\textbf{100.00} &	\textbf{100.00} &	\textbf{100.00} &	\textbf{100.00} \\
		\hline
		\hline
		& \multicolumn{2}{|c}{1/8 R($\%$)} & \multicolumn{2}{|c}{1/4 R($\%$)} & \multicolumn{2}{|c}{1/2 R($\%$)} &  \multicolumn{2}{|c}{2/3 R($\%$)} & \multicolumn{2}{|c|}{1R($\%$)}\\
		\hline
		\texttt{PALM} & Normal & Diseased & Normal & Diseased & Normal & Diseased & Normal & Diseased & Normal & Diseased \\
		\hline
		UNet (2015)~\cite{ronneberger2015u} & 57.45&	9.43&	74.47&	18.87&	76.60&	41.51&	76.60&	50.94&	76.60&	64.15 \\
		U2 Net (2020)~\cite{qin2020u2} & \textit{70.21} &	\textit{11.32} &	\textit{93.62} &	\textit{28.30} &	\textit{95.74} &	\textit{60.38} &	95.74 &	\textit{77.36} &	\textit{97.87} &	\textit{84.91} \\
		TransUNet (2021)~\cite{chen2021transunet} & \textbf{82.98} &	5.66 &	\textbf{95.74} &	18.87 &	\textbf{97.87} &	43.40 &	\textit{97.87} &	52.83 &	\textit{97.87} &	75.47 \\
		Bilateral-ViT (\textbf{Proposed}) & \textbf{82.98} &	\textbf{13.21} &	\textbf{95.74} &	\textbf{37.74} &	\textbf{97.87} &	\textbf{69.81} &	\textbf{100.00} &	\textbf{81.13} &	\textbf{100.00} &	\textbf{92.45} \\
		\hline
	\end{tabular}
\end{table*}

\textbf{Multi-scale Feature Fusion (MFF) blocks}. In contrast to the plain convolutional decoder blocks of the basic TransUNet, we use three Multi-scale Feature Fusion (MFF) blocks as the decoders for effective multi-scale feature fusion.  The input to each MFF block is the concatenation of three types of features: (1) the multi-scale skip-connection features from the main branch, (ii) the hidden feature encoded by the last transformer block or the previous MFF block, (iii) the multi-scale SIG features from the vessel branch. The architecture of the MFF blocks is illustrated in Fig.~\ref{Fig_Arch2}-b, which is similar to one of the SIG blocks. From MFF block\_1 to MFF block\_3, we gradually increase the number of network layers in each MFF block. In this way, the later MFF blocks can capture more spatial context corresponding to larger feature maps. In the end, the concatenated feature maps of MFF block\_3 and SIG block\_4 are passed to two convolutional layers for outputting the fovea region score maps.

\subsection{Implementation Details}
We first remove the uninformative black background from the original fundus image, then pad and resize the cropped image region to a spatial resolution of $512\times512$. We perform intensity normalization and data augmentation on the input images of the main branch and the vessel branch. To train our Bilateral-ViT network, we generate circular fovea segmentation masks from the ground-truth fovea coordinates. During the testing phase, we apply the sigmoid function to network prediction for the probabilistic map. We then collect all pixels with significant probabilistic scores and calculate their median coordinates as the final fovea location coordinates.

All experiments were coded using PyTorch and conducted on one NVIDIA GeForce RTX TITAN GPU. The weights of convolutional and linear layers were initialized by Kaiming initialization protocol~\cite{he2016identity}. The initial learning rate was $1e^{-3}$ which gradually decays to $1e^{-7}$ over 200 epochs using the Cosine Annealing LR strategy. The optimizer was Adam~\cite{kingma2014adam} and the batch size 2. We employed a combination of dice loss and binary cross-entropy as the loss function.

\section{Experiments}
\label{sec:exp}

We performed experiments using the \texttt{Messidor}~\cite{decenciere2014feedback} and \texttt{PALM}~\cite{55pk-8z03-19} datasets. The \texttt{Messidor} dataset is for diabetic retinopathy analysis. It consists of 540 normal and 660 diseased retinas.  We utilized 1136 images from this dataset with fovea locations provided by~\cite{gegundez2013locating}. The \texttt{PALM} dataset was released for the Pathologic Myopia Challenge (PALM) 2019.  It consists of 400 images annotated with fovea locations, in which 213 images are pathologic myopia, and the remaining 187 images are normal retinas. For fairness of comparison, we keep our data split identical to~\cite{xie2020end}. 

To evaluate the performance of fovea localization,  we adopt the following evaluation protocol~\cite{gegundez2013locating}: the fovea localization is considered successful when the Euclidean distance between the ground-truth and predicted fovea coordinates is no larger than a predefined threshold value, such as the optic disc radius $R$. For a comprehensive evaluation, accuracy corresponding to different evaluation thresholds (for example, $2 R$ indicating the predefined threshold values are set to twice the optic disc radius $R$) is usually reported.

\subsection{Fovea Localization on Normal and Diseased Images}
\begin{figure*}
	\centering
	\includegraphics[width=5in]{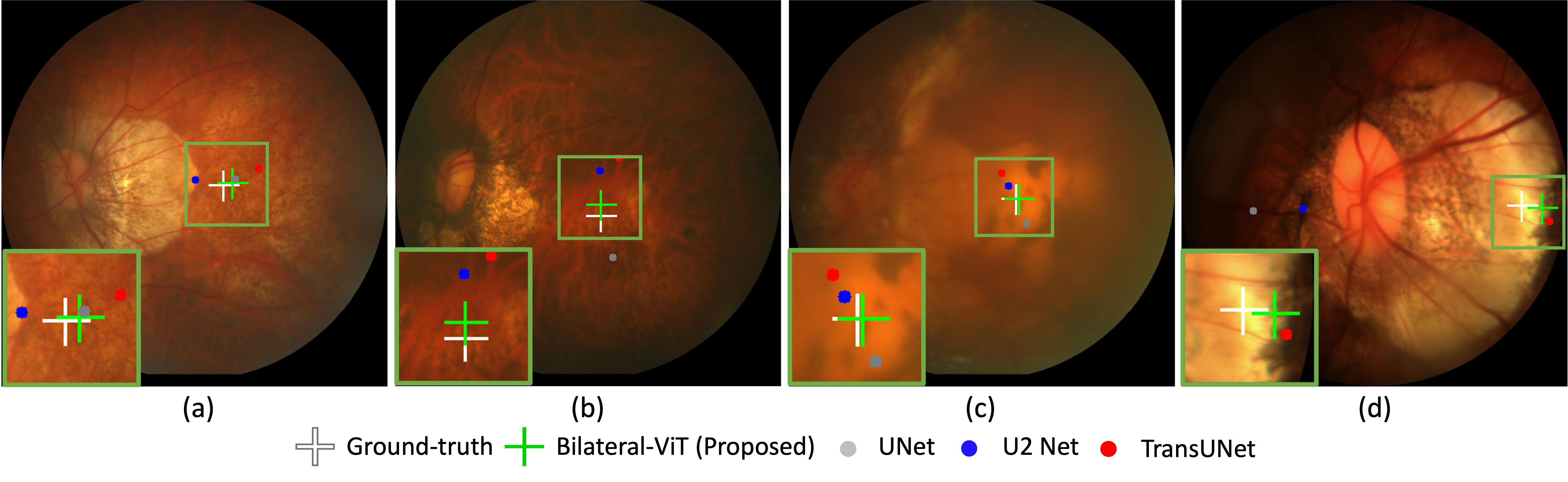}
	\caption{\footnotesize Visual results of fovea localization predicted by different methods.} \label{Fig_1}
\end{figure*}

In Table~\ref{table-path}, we evaluate the performance of normal and diseased cases separately. We reimplement several widely used segmentation networks as comparison baselines, such as UNet~\cite{ronneberger2015u}, U2 Net~\cite{qin2020u2}, and TransUNet~\cite{chen2021transunet}. Bilateral-ViT obtains 100\% accuracy from $1/2 R$ to $1R$ on all the \texttt{Messidor} images, and 100\% accuracy from $2/3 R$ to $1R$ on the normal \texttt{PALM} images. Thus demonstrating that the performance of Bilateral-ViT is highly reliable for normal fundus images.

For the diseased cases in the \texttt{PALM} dataset, Bilateral-ViT reaches 92.45\% foveal localization accuracy for the threshold of $1R$ and significantly outperforms the second-best results by a large margin (7.54\%). Fig.~\ref{Fig_1} provides some visual results of fovea localization on diseases images from the \texttt{PALM} dataset.  Our Bilateral-ViT generates the most accurate predictions for the severely diseased images with large atrophic regions (see Fig.~\ref{Fig_1}-a and Fig.~\ref{Fig_1}-b), or the heavily blurred image (see Fig.~\ref{Fig_1}-c). In Fig.~\ref{Fig_1}-d where the fovea is close to the image border, the predicted fovea locations from baseline networks (UNet and U2 Net) appear on the wrong side of the optic disc. However, TransUNet~\cite{chen2021transunet} and our method still perform well, potentially due to their long-range modeling capability. Such results highlight that our proposed Bilateral-ViT has a significant advantage for diseased cases.

\subsection{Comparison with State-of-the-Art Methods}
\begin{table}\tiny
	\caption{\footnotesize Comparison with existing studies using the \texttt{Messidor} and \texttt{PALM} datasets based on the $R$ rule. The best and second best results are highlighted in bold and italics respectively.}\label{table-SOTA}
	\centering
	\begin{tabular}{|l |c |c |c |c |c |}
		\hline
		\texttt{Messidor} & 1/8 R (\%) & 1/4 R (\%) & 1/2 R (\%) & 1R (\%) & 2R (\%) \\
		\hline
		Gegundez-Arias \etal (2013)~\cite{gegundez2013locating}  & - & 76.32 & 93.84 & 98.24 & 99.30 \\
		Aquino (2014)~\cite{aquino2014establishing} 		 	& - 		& 83.01 	& 91.28 	& 98.24 	& 99.56 \\
		Dashtbozorg \etal (2016)~\cite{dashtbozorg2016automatic}  	& - 		& 66.50 	& 93.75	& 98.87	& - \\
		Girard \etal (2016)~\cite{girard2016simultaneous} 	& - 		& - 		& 94.00 	& 98.00 	& - \\
		Molina-Casado \etal (2017)~\cite{molina2017fast}	& - 		& - 		& 96.08 	& 98.58 	& 99.50 \\
		Al-Bander \etal (2018)~\cite{al2018multiscale}  	& - 		& 66.80 	& 91.40 	& 96.60 	& 99.50 \\
		Meyer \etal (2018)~\cite{meyer2018pixel}	 	& 70.33 	& 94.01 	& 97.71 	& 99.74 	& - \\
		GeethaRamani \etal (2018)~\cite{geetharamani2018macula} & - 		& 85.00 	& 94.08 	& 99.33	& - \\
		Zheng \etal (2019)~\cite{zheng2019new} 	& 60.39	& 91.36 	& 98.32	& 99.03	& - \\
		Huang \etal (2020)~\cite{huang2020efficient}  & - 		& 70.10 	& 89.20 	& 99.25 	& - \\ 
		Xie \etal (2020)~\cite{xie2020end} & \textit{83.81} 	& \textit{98.15} 	& \textit{99.74} 	& \textit{99.82} 	& \textbf{100.00} \\
		Bilateral-ViT (\textbf{Proposed}) &  \textbf{85.65} & \textbf{98.59} & \textbf{100.00} & \textbf{100.00} & \textbf{100.00} \\
		\hline
		\hline
		\texttt{PALM} & 1/8 R (\%) & 1/4 R (\%) & 1/2 R (\%) & 2/3 R (\%) & 1R (\%) \\
		\hline
		Xie \etal (2020)~\cite{xie2020end} 	 	& - 	& - 	& - 	& \textit{87}	& \textit{94} \\
		Bilateral-ViT (\textbf{Proposed}) & \textbf{46} & \textbf{65} & \textbf{83} & \textbf{90} & \textbf{96} \\
		\hline
	\end{tabular}
\end{table}

From Table~\ref{table-SOTA}, the Bilateral-ViT achieves state-of-the-art performance for all the evaluation settings. In particular, on the \texttt{Messidor} dataset, at $1/8 R$, our network reaches the best accuracy of 85.65\% with a gain of 1.84\% compared to the second-best score (83.81\%)~\cite{xie2020end}. It also reaches an accuracy of 100\% at evaluation thresholds of $1/2 R$, $1 R$, and $2 R$; in other words, the localization errors are at most $1/2 R$ (approximately 19 pixels for an input image size of  $512\times512$). \texttt{PALM} is a considerably more challenging dataset due to fewer images and complex diseased patterns. Our method achieved accuracies of 90\% and 96\% at $2/3 R$ and $1R$, which is 3\% and 2\% better than the previous work~\cite{xie2020end}, respectively.

\begin{table}\tiny
	\caption{\footnotesize \textbf{Top} and \textbf{Bottom}: Performance of the ablation study using the \texttt{Messidor} and \texttt{PALM} datasets respectively. VB refers to the vessel branch. The best and second best results are highlighted in bold and italics.}\label{table-trans}
	\centering
	\begin{tabular}{|l |c |c |c |c |c |c |c |}
		\hline
		\texttt{Messidor} & 1/8 R (\%) & 1/4 R (\%) & 1/2 R (\%) & 1R (\%) & 2R (\%) \\
		\hline
		ViT+plain decoder (TransUNet~\cite{chen2021transunet}) & \textbf{85.74}&	97.98&	\textit{99.91}&	\textit{99.91}&	\textit{99.91} \\
		ViT+VB+plain decoder&  {85.56} & 	\textit{98.33} & 99.74 & \textit{99.91} & \textit{99.91}  \\
		{ViT+VB+MFF (\textbf{Proposed})} &  \textit{85.65} & \textbf{98.59} & \textbf{100.00} & \textbf{100.00} & \textbf{100.00}   \\
		ViT+VB (fundu as the input)+MFF & \textit{85.65} & 97.89 & \textit{99.91} & \textbf{100.00} & \textbf{100.00}   \\
		\hline
		\hline
		\texttt{PALM} & 1/8 R (\%) & 1/4 R (\%) & 1/2 R (\%) & 2/3 R (\%) & 1R (\%) \\
		\hline
		ViT+plain decoder (TransUNet~\cite{chen2021transunet}) & 42 &	55&	69&	74&	86 \\
		ViT+VB+plain decoder & \textit{45} & 52 & 72 & 77 & 85 \\
		{ViT+VB+MFF (\textbf{Proposed})} & \textbf{46} & \textbf{65} & \textbf{83} & \textbf{90} & \textbf{96} \\
		ViT+VB (fundu as the input)+MFF & {43} & 	\textit{58} & \textit{82} & \textit{89} & \textbf{96} \\
		\hline
	\end{tabular}\\
\end{table}

\subsection{Ablation Study and Cross-Dataset Experiments}
\label{sec:ablation}
We conducted a comprehensive set of ablation experiments to evaluate the effectiveness of different components (see Table~\ref{table-trans}):
\begin{itemize}
	\item ViT+plain decoder: the TransUNet architecture~\cite{chen2021transunet} comprised of a vision transformer-based encoder and a plain decoder used as the comparison baseline. 
	\item ViT+VB+plain decoder: we add the vessel branch (vessel segmentation mask as the input) to the baseline network.
	\item ViT+VB+MFF (\textbf{the proposed Bilateral-ViT}): we add the vessel branch (vessel segmentation mask as the input) and MFF blocks to the baseline network.
	\item ViT+VB (fundus as the input)+MFF: we add the vessel branch (fundu image as the input) and MFF blocks to the baseline network. This configuration compares the performance differences between fundus images and vessel segmentation maps as inputs to the vessel branch. 
\end{itemize}

The performance of ``ViT+plain decoder (TransUNet)'' and ``ViT+VB+plain decoder'' are similar on both datasets; a possible reason is that the plain decoder does not have adequate capacity to fuse features from the vessel branch and transformer blocks. By further adding MFF blocks, the proposed Bilateral-ViT (ViT+VB+MFF) shows superior performance, suggesting the significance of the customized MFF blocks. The performance of ``ViT+VB+MFF' is much better than ``ViT+VB (fundus as the input)+MFF'', demonstrating the usefulness of the vessel segmentation map.  On the other hand, we note that ``ViT+VB (fundus as the input)+MFF'' outperforms all the existing works, implying our network can achieve the state-of-the-art performance even without the input of a vessel segmentation map.

\begin{table}\tiny
	\caption{\footnotesize Performance of cross-dataset experiments. The models used here are exactly those in the \textbf{Bottom} of Table~\ref{table-trans}. They were constructed using \texttt{PALM} only and generated the following results on \texttt{Messidor}. The higher results based on the $R$, and the lower results based on distance errors, are better. VB refers to the vessel branch. The best and second best results are highlighted in bold and italics respectively.}\label{table-cross}
	\centering
	\begin{tabular}{|l |c |c |c |c |c |c |c |c lc |}
		\hline
		Cross-Dataset & 1/8 R(\%) & 1/4 R(\%) & 1/2 R(\%) & 1R(\%) & 2R(\%) & Errors\\
		\hline
		Xie \etal~\cite{xie2020end} & - & - & - & 95.26 & - & 22.84 \\
		ViT+plain decoder (TransUNet) & 77.82&	\textit{95.95}&	\textbf{98.59}&	\textit{99.03}&	99.30& 10.76 \\
		ViT+VB+plain decoder & \textit{78.17} & {95.69} & {98.24} & {98.77} & 99.12 & 11.38 \\
		{ViT+VB+MFF (\textbf{Proposed})} & \textbf{81.78} & \textbf{96.48} & \textit{98.42} & \textbf{99.38} & \textbf{100.00} & \textbf{8.57} \\
		ViT+VB (fundu as the input)+MFF & 77.02 & 94.28 & 97.62 & 98.68 & \textit{99.47} & \textit{10.69} \\
		\hline
	\end{tabular}
\end{table}

We conducted cross-dataset experiments to assess the generalization capability of the proposed Bilateral-ViT. The models were trained on the \texttt{PALM} dataset and tested on the \texttt{Messidor} dataset. From Table~\ref{table-cross}, the accuracy is 99.38\% at $1R$, which is a 4.12\% improvement over the best-reported result (95.26\%). The average localization error for the original image resolution is 8.57 pixels compared to the previous best result of 22.84 pixels. In addition, the proposed Bilateral-ViT outperforms the baselines by a significant margin, especially for $1/8 R$, demonstrating its robustness for the cross-dataset setting.

\section{Conclusions}
\label{sec:con}
This paper proposes a novel Vision Transformer (ViT) approach for robust fovea localization. It consists of a transformer-based main network branch for integrating global context and a vessel branch for explicitly incorporating the structure of blood vessels. The encoded features are subsequently merged with a customized Multi-scale Feature Fusion (MFF) module. Our experiments demonstrate that the proposed approach has a significant advantage in handling diseased images. It also has excellent generalization capability, as shown in the cross-dataset experiments.  Thanks to the transformer-based feature encoder, the incorporation of blood vessel structure, and the carefully designed MFF module, our approach establishes the new state of the arts on both \texttt{Messidor} and \texttt{PALM} datasets.


\section{Compliance with Ethical Standards}
\label{sec:con}

This research study was conducted retrospectively using human subject data available in the open access \texttt{Messidor} and \texttt{PALM} datasets. Ethical approval was not required as confirmed by the license attached with the open access data.
\section*{Acknowledgment}

This work was supported in part by the National Science Foundation of China (NSFC) under Grant 61501380, in part by the Key Program Special Fund in XJTLU (KSF-A-22), in part by the Neusoft Corporation (item number SKLSAOP1702), in part by Voxelcloud Inc.



%



\bibliographystyle{IEEEtran}
\bibliography{Manuscript}

\end{document}